\documentclass[pre]{revtex4-2}
\usepackage[utf8]{inputenc}
\usepackage{graphicx}
\usepackage{amsmath}
\usepackage{amssymb}
\usepackage{color}
\usepackage{todonotes}

\begin{document}
\title{Indentation of an elastic disk on a circular supporting ring}

\author{Tristan Suzanne, Julien Deschamps, Marc Georgelin, Gwenn Boedec}
\affiliation{Aix Marseille University, CNRS, Centrale Marseille, IRPHE,
Marseille, France}

\begin{abstract}
Thin elastic two-dimensionnal systems under compressive stresses may relieve part of their stretching energy by developing out of plane undulations. We investigate experimentally and theoretically the indentation of an elastic disk supported by a circular ring and show that compressive stresses are relieved via two different routes : either developing \textit{buckles} which are spread over the system or developing a \textit{d-cone} where deformation is concentrated in a subregion of the system. We characterize the indentation threshold for wrinkles or d-cone existence as a function of aspect ratio.
\end{abstract}

\date{\today}

\maketitle

\section{Introduction}

Indentation of an elastic material through a circular hole occurs in many situations, like in metal forming processes such as deep drawing \cite{Schey} where a sheet of metal is pushed with a punch through a cavity, or to probe the mechanical properties of two-dimensional materials \cite{Cao_2019} such as graphene where suspended layers can be deformed with AFM tips. In both macroscale deep-drawing experiments on metal sheets and atomic-thin films indentation tests, a characteristic wrinkling instability may appear \cite{Yu-1981-the-buckling,Yu-Zhang-1986-the-elastic,Dai_2021, Davidovitch_2021}. Indeed, pushing an elastic disk through a hole induces compressive orthoradial stresses, which may be relieved by out-of-plane undulations called wrinkles. The onset of wrinkling on circular elastic sheets, their extent and their wavelength have been the subjects of numerous works in various situations, where the compressive stresses are generated by shearing \cite{Miyamura_2000}, capillary forces \cite{Huang-2007-Capillary}, suction \cite{Geminard_2004}, indentation of floating sheets \cite{Box_2017} or gravity-induced deflection \cite{Delapierre-2018-Wrinkling}. A prototypical situation to study wrinkles apparition in these situations is the Lam\'e setup \cite{Davidovitch_2011} where an initially flat annulus of elastic material is subjected to differential tension at the inner and the outer edge. However, this setup is not always realized in practice, because significant deflection of the annulus may occur prior to wrinkling, changing the pre-wrinkled stress field. Moreover, the importance of the aspect ratio of the system has been highlighted in theoretical works \cite{Coman-2006-On-some,Coman_2007} and in experiments on floating annulus stretched by differential surface tension \cite{Pineirua_2013}, yet most studies focus on the limit of infinite sheets. 

On the other hand, thin elastic sheets also show a limit behavior called "inextensible" where, if possible, all the elastic energy is stored as bending energy and stretching deformation is avoided : in the indentation process, this is prohibited by Gauss' \emph{Theorema Egregium}, which implies that passing from a flat state (zero Gaussian curvature) to a doubly curved state (non zero Gaussian curvature) cannot be isometric everywhere in the sheet. This leads to the formation of the so called \emph{d-cone} \cite{BenAmar_1997,Cerda-prl1998}, where deformation is isometric almost everywhere, except for a stretching core under the indenter tip. The d-cone is also a building block for the description of crumpling \cite{Witten-2007-stress-focusing-in}, and recent works have shown that its particular mechanical properties may be suitable for generating electricity via flexoelectricity \cite{Kodali_2017,Wang_2019,Javvaji_2021}. The mechanics of d-cone has been studied theoretically \cite{BenAmar_1997,Cerda-prl1998,Cerda_2005} and experimentally \cite{Chaieb_prl1998,Cerda_nat1999}, with a particular focus on the force-indentation relationship as well as the apparition of crescent shapes in the core region, with a scaling of the size of the core depending on thickness and indentation \cite{Cerda_2005}. However, in these experiments, a wrinkled state has never been reported while, as discussed in the preceeding paragraph, it can be expected from experiments on similar systems. Indeed, the indentation of an elastic disk through a circular hole is a system that may exhibit two different ways of relieving the stretching energy : either wrinkling where some stretching is still present in a diffuse manner, or by forming a d-cone, where the stretching energy is focused in a small zone. To our knowledge, the connection between the two different responses of the system has not been investigated.

In this paper, we conduct experiments on various elastic materials to obtain a state diagram showing the zone of existence of wrinkles and d-cone as a function of the aspect ratio and the central indentation. We show that both require to reach a (different) critical indentation to appear, which depends on the thickness of the membrane and on the aspect ratio. We rationalize these experimental findings for wrinkling by conducting a linear stability analysis of the system, which is in good agreement with the wrinkling boundary, and discuss the threshold for d-cone with a scaling analysis of the energies involved.

\section{System description}
\begin{figure}[!h]
\begin{center}
\includegraphics[scale=0.35]{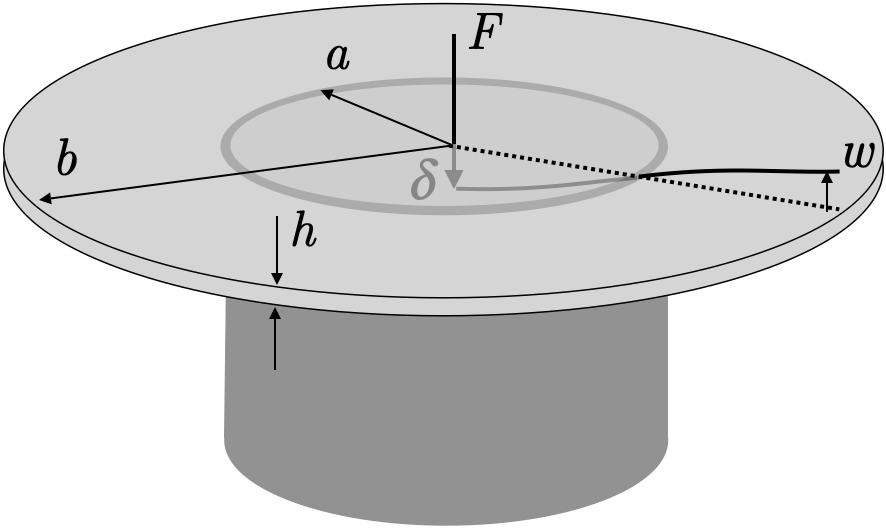}
\end{center}
\caption{A disk of elastic material of radius $b$ and thickness $h$ is deposited on top of a circular support of radius $a$ and indented of $\delta$ with a tip exerting a force $F$ at the center. The deflection of the initially flat disk is noted $w$.}
\label{fig:descriptionsyst}
\end{figure}

We consider an initially flat disk of radius $b$ and uniform thickness $h$ such that $h/b \ll 1$. The disk is placed on a circular ring of radius $a$ and indented in its center. The indentation depth and force are $\delta$ and $F$ respectively (figure \ref{fig:descriptionsyst}). The system is thus considered as the combination of two sub-regions, the inner part defined by $0\leq r < a$ and the outer annulus defined by $a <  r \leq b$. These two sub-regions are coupled by boundary conditions at the supporting ring location.\\ 

\begin{figure}[!h]
\includegraphics[scale=0.15]{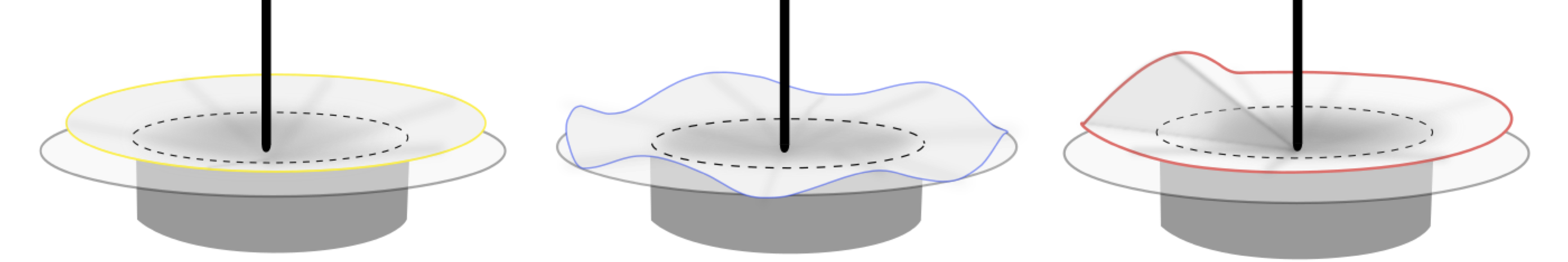}
\caption{Three different states of deformation can be observed during the indentation process. Left : axisymmetric state. Center : buckled state. Right : d-cone state.}
\label{fig:descriptionstates}
\end{figure}

As the indentation depth increases, the disk may experience three different states (see figure \ref{fig:descriptionstates}), axisymmetric, buckled and d-cone shapes, switching from one to another by successive instabilities. For small indentations, the whole disk deflects upwards axisymmetrically. It adopts a curved shape revealing a non-zero gaussian curvature accompanied by a compressive azimuthal stress : this is the axisymmetric state. 
Increasing again the indentation may lead the disk to deform into two different states. A part of the stretching energy accumulated in the axisymmetric state may be relaxed to bending energy which leads to out-of-plane undulations of the outer annulus : this is the buckled state. Or the stretching energy may be relaxed by focusing in a core region near the indentation tip while the rest of the disk adopts a developable conical shape : this is the so-called d-cone. A remarkable difference between the d-cone and the buckled state is that in the buckled state we do not observe a noticeable lift of the disk from the supporting ring, while in the d-cone state there is a clear loss of contact with the supporting ring over an angular portion of $110 \pm 5 ^\circ $\cite{Chaieb_prl1998}.

\section{Methods}
\subsection{Experimental setup}

\begin{figure}
\begin{center}
\includegraphics[scale=0.35]{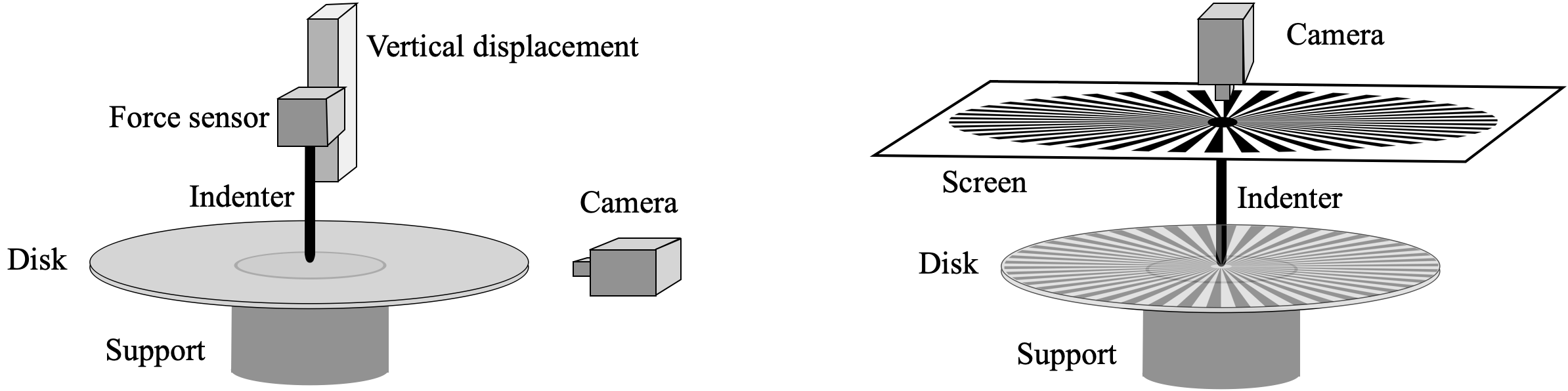}
\end{center}
\caption{Experimental setup. Left : the indenter is fixed on a load cell, which is moved with a translation stage. A camera is used to take side views of the deformation. Right : the camera is used to take top views of the reflection of a pattern in the elastic disk acting as a mirror.}
\label{fig:descriptionXP}
\end{figure}

We developed an experimental setup that allows to vertically indent an horizontal elastic disk supported on a circular ring.
The indenter is made from long screw whose tip is a half sphere of radius 1 mm. The indenter is attached to a load cell to measure the force exerted by the tip. Depending on the disk material, K3D40 50 N from ME-Systeme or 1 N to 8 N load cells from Phidgets are used. The load cell is vertically moved by a precision linear translation stage LMS 180 supplied by Physik Instrumente. The stage and the load cell are controlled by Labview from National Instrument.
As supporting rings we used hollow punches supplied by BOEHM. Additional manual precision stages enable to center and align the indenter with the support ring. A digital camera NIKON 5300 controlled by DigiCamControl software is used to capture picture of the disk (see figure \ref{fig:descriptionXP}(a)) . Image processing is performed with Matlab.\\

\begin{table}[!h]

    \begin{tabular}{c|c|c|c}

  Material & h (mm) & b (mm)  &  E (MPa) \\
  \hline
PET (Mylar) &  \textbf{0.08}($\pm0.01$) - \textbf{0.34}($\pm0.02$) & \textbf{30}($\pm1$) - \textbf{65}($\pm2$) & \textbf{4200}($\pm 200$) \\
\hline
Silicone & \textbf{0.7}($\pm0.1$) - \textbf{1.5}($\pm0.1$) & \textbf{16}($\pm1$) - \textbf{62}($\pm2$) & \textbf{0.5}$(\pm 0.2$) - \textbf{5}($\pm 0.5$) \\
\hline
PDMS & \textbf{0.63}($\pm0.05$) - \textbf{1.4}($\pm0.1$) & \textbf{25}($\pm1$) - \textbf{50}($\pm1$) & \textbf{1.5}$\pm(0.2)$ - \textbf{2.3}($\pm 0.5$) \\
    \end{tabular}
    \caption[Table of materials properties]{Material properties of elastic disks used in this study. Typical uncertainties are indicated in parenthesis.}
    \label{tableau-propriete-exp}
\end{table}

Disks are cut into elastic sheets made of three different materials : PDMS, Silicone and PET (Mylar). PET disks are obtained from standard stencil sheets.
PDMS and Silicone sheets are both made by mixing curing agent and a polymer base. The curing agent concentration was kept constant for silicone RTV 181 purchased from Esprit Composite while for PDMS, Silgard 184 purchased from Merck, we varied the curing agent concentration in order to change the Young modulus by an order of magnitude. Air bubbles are removed from the mixture liquid phase in a vacuum chamber. Then we let the mixture cure for at least 24h at room temperature between two planar parallel plates made of aluminium with a controlled spacing. When unmoulded, we control the homogeneity of the sheet by measuring the thickness in different locations with a Palmer micrometer. This process is repeated to obtain an average value of its thickness (typical variations are reported in table \ref{tableau-propriete-exp}). The Young modulus is measured with a homemade tensile test machine that consists of two clamps that hold the material specimen. One clamp is fixed while the other is linked to a load cell attached to the translation stage. For each material, diameter, thickness and Young modulus ranges are reported in table \ref{tableau-propriete-exp}. 
Once the disk is cut from the sheet and characterized, it is gently deposited and centered on the support ring of radius $a$.

\subsection{Axisymmetric-Buckling transition}
\label{sec:xp_axi_buckling_transition}

\begin{figure}
    \centering
    \includegraphics[scale=0.5]{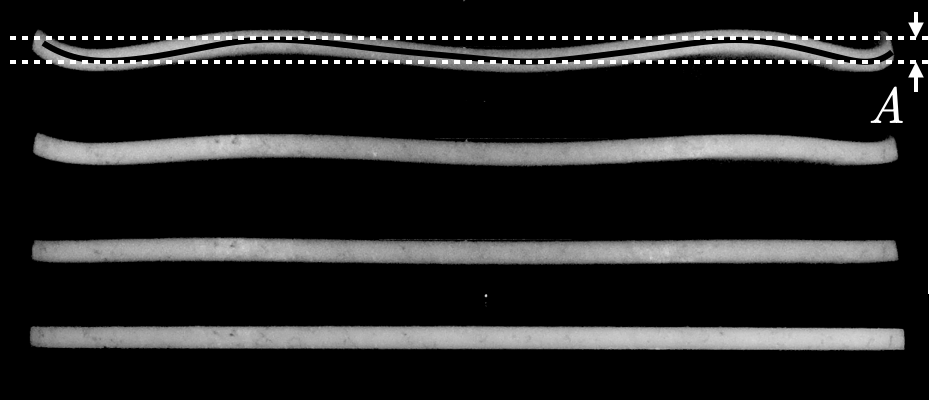}
    \caption{Typical side-view pictures and definition of undulations amplitude $A$ for a silicone disk : $b$=20 mm, $a=$13 mm and $h$=1 mm. From bottom to top : $\delta$=0 mm, $\delta$=4 mm, $\delta$=4.5 mm and $\delta$=5 mm. }
    \label{fig:three-quarter-and-side-view}
\end{figure}

\begin{figure}

    \begin{center}
    \includegraphics[scale=0.5]{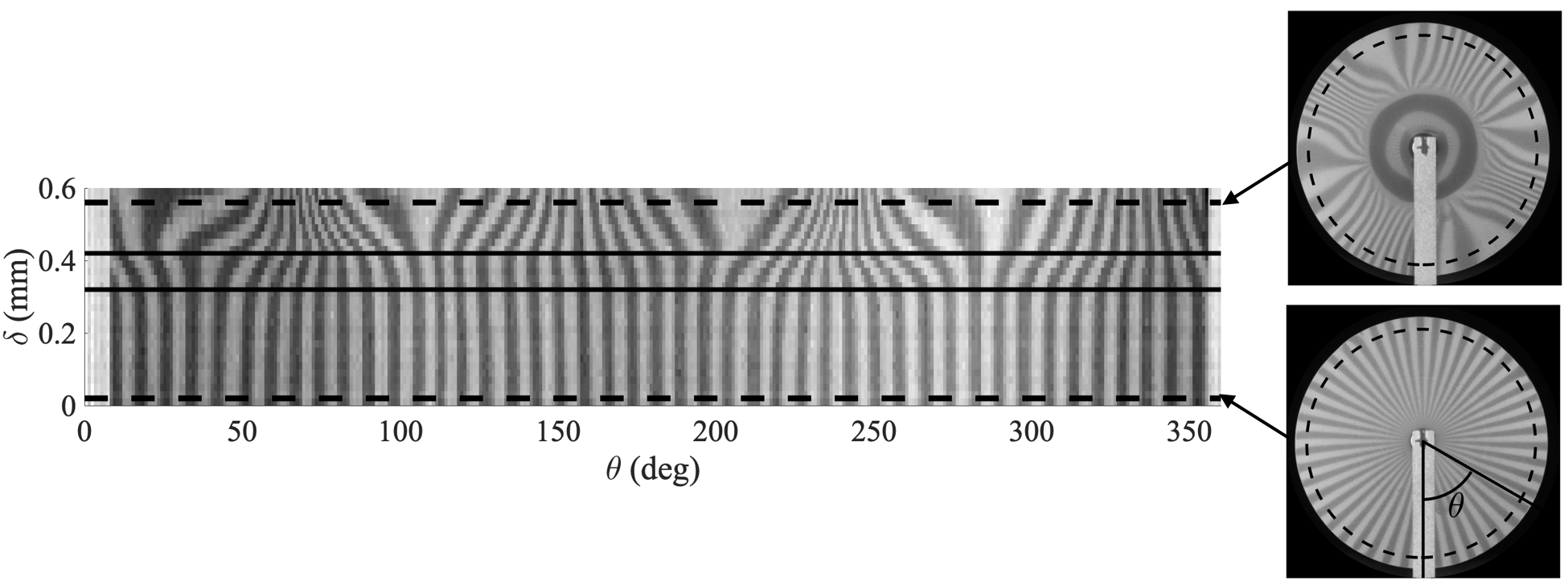}

\caption{Typical angle-indentation diagram used to determine axisymmetric-buckled threshold (see text for the description of the method). Measurements of the deflection is performed on circle of diameter $r=29$ mm (dashed circles on right images) Solid lines represent the lower and upper bounds for the critical indentation. $a=15$ mm, $b=30$ mm, $h=125\ \mu$m. }
    \label{fig:angle-indentation-diagramm}
    \end{center}
\end{figure}
As the indentation increases the disk adopts an increasingly pointed conical shape. Then the edge of the disk remains flat and moves upward. For a critical indentation $\delta_b$ the edge starts to undulate revealing a wrinkling instability. To determine precisely this critical value we observe the disk from a side view and we measure, after image processing on Matlab, the amplitude of the edge undulations as a function of the indentation. A zero amplitude characterizes the axisymmetrical state while a non-zero amplitude reveals the wrinkled state.
This method is appropriate for disks with a millimetric thickness, as our disks made of PDMS and silicone (figure \ref{fig:three-quarter-and-side-view}), but is not suitable for thin disk of Mylar since the undulations could not be properly resolved. For these disks, we use an alternative method based on deflectometry (see figure \ref{fig:descriptionXP}(b)): a screen with a radial pattern of alternating black and white stripes is placed above the flat state disk in a parallel way. Using a small hole in the screen, we take pictures of the reflection of this pattern in the Mylar sheet which acts as a mirror. For axisymmetric deformations, the image shows radial symmetry, while for wrinkled state, this symmetry is broken. To determine wrinkling threshold unambiguously, we construct a diagram of pixel intensity over a circle with angle as abscissa, indentation as ordinate : axisymmetric regime corresponds to indentations where the boundaries between black and white stripes remain straight, while wrinkling apparition is characterized by an apparent convergence of stripes in the convex zones and an apparent divergence of stripes in the concave zones (see figure \ref{fig:angle-indentation-diagramm}). 
From this diagram we measure the orthoradial displacement of the boundaries between black and white radial
stripes as a function of the indentation. This displacement can be related to surface slope using simple geometrical optics relations \cite{Balzer_2010}. We then integrate the slope to obtain the orthoradial deflection. We finally determine a lower and upper bound for the transition by measuring the peak-to-peak amplitude of the orthoradial undulations. The upper bound corresponds to an amplitude $A$ of the disk of about its thickness $h$ as for the lateral observation for thick disks.

\subsection{Stretching - d-cone transition}
\label{sec:methods-stretch-dcone-transition}

\begin{figure}[!h]
    \centering
    \includegraphics[scale=0.5]{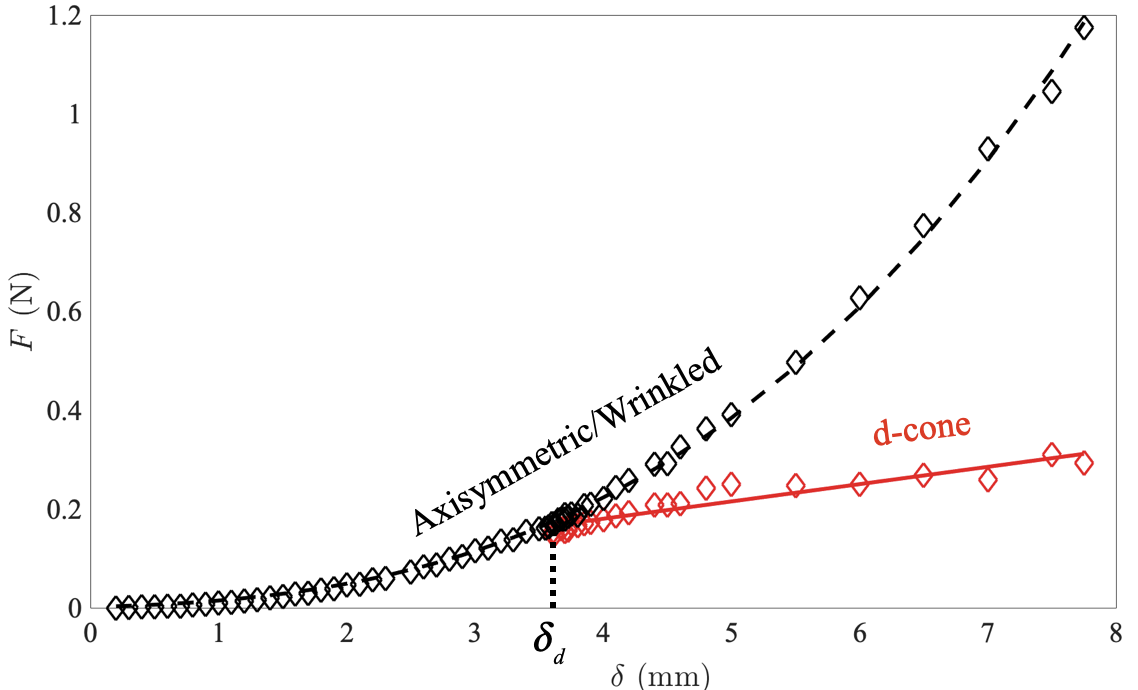}
   \caption{Typical experimental force-indentation curve showing two different branches : a diffuse stretching branch (dashed black curve) and a focused state (solid red curve). On the stretching branch, the system can be either axisymmetric or develop wrinkles, this does not reflect on the force-indentation relationship. Experimental data corresponds to a PDMS disk ($a= 15$ mm, $b=24.5$ mm, $E = 1.5$ MPa)}
    \label{fig:force_indentation}
\end{figure}

As mentioned in the previous sections, the disk can deform into three different states depending on the indentation : axisymmetric, wrinkled and d-cone. Measuring the force-indentation relationship, these three different states corresponds to two different branches (see figure \ref{fig:force_indentation}): the first branch (axisymmetric and wrinkled) is dominated by stretching, giving a force scaling as $\delta^3$, while the second branch (d-cone) is dominated by bending, giving a force scaling as $\delta$. Starting from the flat reference and increasing the indentation, the system follows the stretching branch. At regular indentation intervals, we test the possibility to form a d-cone by slightly raising the outer edge with a tip. For small indentations, the system relax to the stretching branch (monitored by the force value), while for higher indentations the system may find another equilibrium with a smaller force value. Note that, if we do not perturb manually the system to form a d-cone, there is always an indentation such that the stretching branch relaxes spontaneously on the d-cone branch. When the d-cone branch is found, it is possible to switch back to the stretching branch by carefully flattening the d-cone. To determine the intersection $\delta_d$ of the two branches, we then decrease slightly the indentation and check for the existence of two different equilibrium (characterized by different values of force exerted on the indenter). We continue this process until only one branch is found : plotting the force-indentation  curve the critical indentation for the stability of the d-cone is defined as the abscissa of the intersection of the two curves (see figure \ref{fig:force_indentation}).

\section{Model}

\subsection{General hypotheses and outlook of the model}

To analyse the stability of the plate with respect to wrinkling of the outer edge, we proceed in two steps. First, we compute an axisymmetric unwrinkled nonlinear base state by solving the full F\"oppl-von Karman (FvK) equations with appropriate boundary conditions, and then we proceed to a linear stability analysis of this base state. Note that keeping nonlinear terms in the description of the base state is crucial to predict correctly the boundaries of stability.
Stationary axisymmetric FvK equations can be written in a general form by using an Airy function $\phi$ \cite{Audoly_2010}:

\begin{equation}
    \begin{split}
        D \Delta^2 w - [\phi,w] &= q(r)\\
        \Delta^2 \phi + \frac{1}{2} E h [w,w] &= 0
    \end{split}
\end{equation}

with $E$ the Young modulus of the material, $\nu$ its Poisson coefficient, $D=\frac{Eh^3}{12(1-\nu^2)}$ the plate bending modulus, $h$ the plate thickness. The load is defined by $q(r) =- \frac{F \underline{\delta}(r)}{2 \pi r} +\frac{R\delta(r-a)}{2\pi r} - \rho g h$, where $\underline{\delta}(r)$ is the two-dimensional Dirac function written in polar coordinates, $R$ is the reaction force due to supporting ring and $\rho$ the volume mass of the plate. The operator $[f,g]$ is \begin{equation}
\begin{split}
[f,g]=&\frac{\partial^2 f}{\partial r^2}\left(\frac{1}{r}\frac{\partial g}{\partial r}+\frac{1}{r^2}\frac{\partial^2 g}{\partial \theta^2}\right)+\frac{\partial^2 g}{\partial r^2}\left(\frac{1}{r}\frac{\partial f}{\partial r}+\frac{1}{r^2}\frac{\partial^2 f}{\partial \theta^2}\right)\\
&-2\left(\frac{1}{r}\frac{\partial^2 f}{\partial r\partial \theta}-\frac{1}{r^2}\frac{\partial f}{\partial \theta}\right)\left(\frac{1}{r}\frac{\partial^2 g}{\partial r\partial \theta}-\frac{1}{r^2}\frac{\partial g}{\partial \theta}\right)
\end{split}
\end{equation}

The Airy function is related to stresses by: 
\begin{equation}
\begin{split}
&h\sigma_{r} = \frac{1}{r}\frac{\partial \phi}{\partial r}+ \frac{1}{r^2}\frac{\partial^2 \phi}{\partial \theta^2} \quad , \quad h\sigma_{\theta} = \frac{\partial^2 \phi}{\partial r^2} \\
&h\sigma_{r\theta} = - \frac{\partial}{\partial r}\left(\frac{1}{r}\frac{\partial \phi}{\partial \theta}\right)
\end{split}    
\end{equation}
where $\sigma_{r}$, $\sigma_{\theta}$ and $\sigma_{r \theta}$ are the radial, orthoradial and shear stresses respectively. These equations are solved over the interval $r \in[0,b]$, with the following boundary conditions :
\begin{equation}
    \begin{split}
    w_{|r=0}&=\delta\\
    u_{|r=0}&=0 \\
    \left(\frac{\partial w}{\partial r}\right)_{|r=0}&=0 \\
    (Q_r)_{|r=b} &= 0 \\
    (M_r)_{|r=b} &= 0 \\
    (\sigma_{r})_{|r=b} &= 0
    \end{split}
    \label{eq:bc1}
\end{equation}
where $u$ is the radial displacement, $Q_r$ is the transverse shear force and $M_r$ the radial moment. Additionally, the presence of a supporting ring of radius $a$ is accounted for by the constraint 
\begin{equation}
    w_{|r=a} = 0
    \label{eq:bcsup}
\end{equation}

We define non-dimensional variables as:
 \begin{align*}
    \Bar{w}&= \frac{w}{h} & \Bar{r}&=\frac{r}{b}\\
    \Bar{\phi}&=\phi \ \frac{12(1-\nu^2) }{Eh^{3}} & G &= \frac{\rho g  b^4}{D} \\
    \end{align*}
With this choice of variables, FvK equations read :
\begin{equation}
    \begin{split}
        \Delta^2 \bar{w} - [\bar{\phi},\bar{w}] &= \bar{q}(\bar{r})\\
        \Delta^2 \bar{\phi} + 6 (1-\nu^2)[\bar{w},\bar{w}] &= 0
    \end{split}
\end{equation}
For the sake of clarity, we drop the bar over dimensionless variables in the remainder of the manuscript, and we explore the stability of axisymmetric deformations as a function of the following dimensionless parameters : $\alpha=a/b$ the aspect ratio, $\delta$ the dimensionless indentation. In the following, we will focus on situations where gravity is negligible, and thus take $G=0$ for the remainder of the analysis.

\subsection{Axisymmetric base state}

We compute axisymmetric solutions of FvK equations by numerical integration using \texttt{bvp4c} function in \texttt{MATLAB}. Because the indentation force is modeled as punctual and located at $r=0$, it is easier for numerical treatment to first integrate analytically the equations once, leading to:

\begin{equation}
    \begin{split}
        &\frac{\partial }{\partial r} \Delta w - \frac{1}{r}\frac{\partial \phi}{\partial r}\frac{\partial w}{\partial r}= Q(r) \\
        &r \frac{\partial}{\partial r}\Delta \phi + 6(1-\nu^2) \left(\frac{\partial w}{\partial r}\right)^2=0
    \end{split}
    \label{eq:fvk_integrated}
\end{equation}

with 

\begin{equation}
        Q(r) = \textcolor{red}{-}\frac{b^2 F}{Dh 2 \pi r} \quad \text{for} \quad  r\in[0,\alpha[ \quad ; \quad 0 \quad  \text{otherwise}
\end{equation}

Note that the load is discontinuous at $r=\alpha$ because of the presence of the support which induces an additional ring of upward force (necessary to balance the indenter force).

\begin{figure}
        \includegraphics[width=0.95\textwidth]{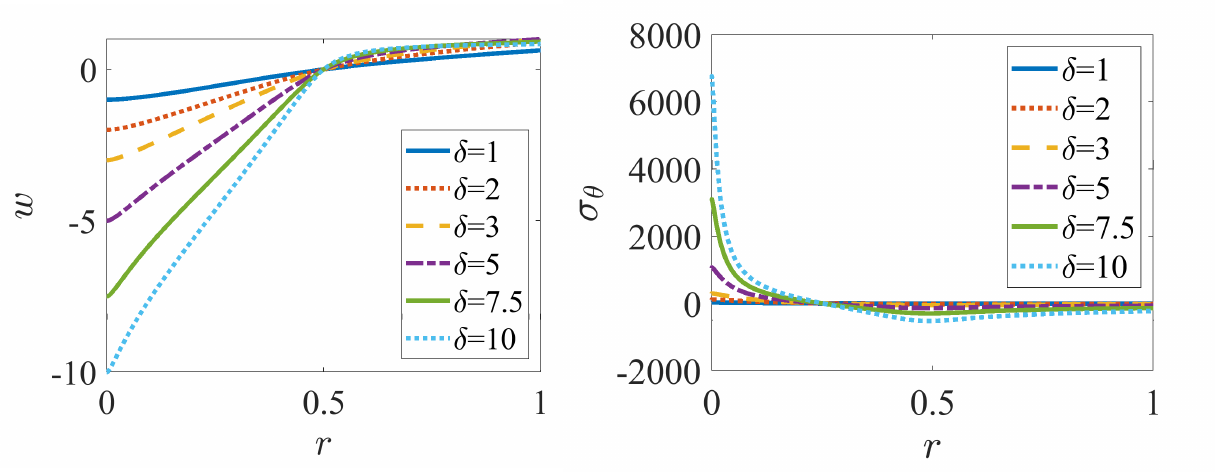} 
    \caption{($G=0,\alpha=0.5$) Left : $w(r)$ profiles for different values of the indentation. Right : Corresponding orthoradial stress profiles.}
    \label{fig:axi_profiles}
\end{figure}

\begin{figure}
        \includegraphics[width=0.95\textwidth]{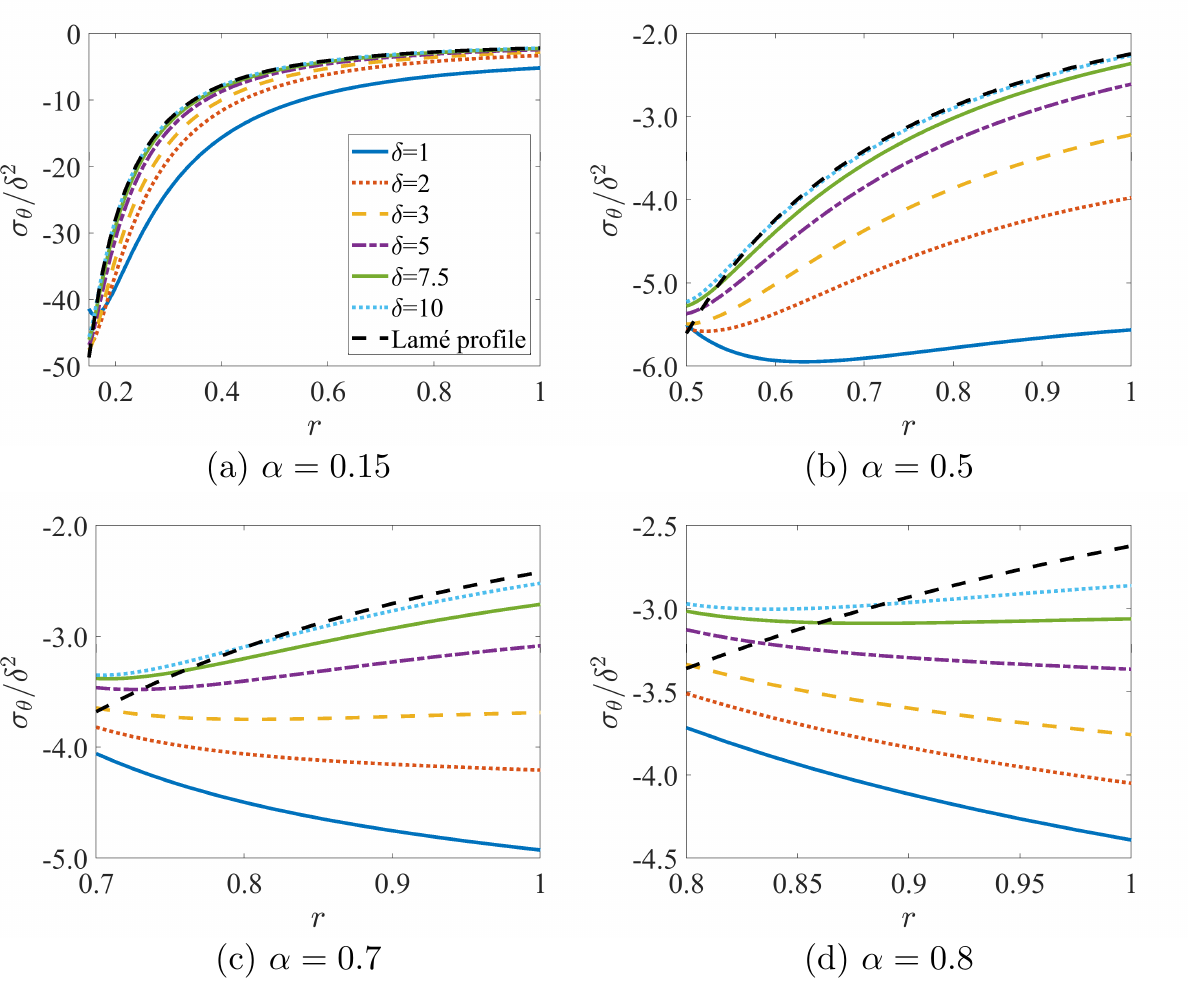} 
 \caption{Orthoradial stress rescaled by $\delta^2$(only the part corresponding to the outer annulus is shown). In the strong indentation limit, the stress collapse to the Lam\'e profile, which we compute analytically using the value of radial stress on the support as a matching parameter.}
    \label{fig:stressProfileRescaled}
    \end{figure}

The corresponding profiles for negligible gravity and intermediate aspect ratio are shown in figure \ref{fig:axi_profiles} (a) : in this case, the inner part of the disk adopts an almost conical shape, while the outer part is also deflected but, as the indentation increases, this outer part adopts an almost flat shape, due to the cost of circumferential bending energy. The orthoradial stresses, whose evolution is shown in figure \ref{fig:axi_profiles} (b) are compressive ($\sigma_\theta<0$) close to the support and in the whole outer part. When the indentation is strong enough to reach the flat outer annulus regime, the orthoradial stresses collapse to the Lam\'e profile, as shown in figure \ref{fig:stressProfileRescaled} (a) and (b). The Lam\'e solution in the outer part writes $\sigma_\theta(r) = \sigma_\alpha \left(1-\frac{1}{1-\alpha^2}(1+\left(\frac{\alpha}{r}\right)^2)\right)$ with $\sigma_\alpha$ the value of the radial stress on the support. Note that this value is not known a priori and results from the matching of the inner part and the outer part of the disk: we determine it from the numerical integration of equations (\ref{eq:fvk_integrated}), as to our knowledge, no analytical solution of the inner part is available. When the aspect ratio $\alpha$ approaches $1$, the annulus is not flat and thus the Lam\'e solution does not describe adequately the stress evolution, as can be seen on figure \ref{fig:stressProfileRescaled} (c) and (d).

\subsection{Stability analysis}
\label{sec:stability_method}

To analyse the stability of the axisymmetric state, we write $w=w_{axi}+w_{pert}$ with $w_{pert}(r,\theta)=w_n(r) e^{\mathrm{i}n\theta}$, and a similar decomposition for $\phi$. The linearisation of FvK equations around the axisymmetric state leads to the following equations, with $\lambda$ the eigenvalue of the buckling mode $(w_{pert},\phi_{pert})$ \cite{Delapierre-2018-Wrinkling}.

\begin{equation}
    \begin{split}
        -\lambda w_{pert} + \Delta^2 w_{pert} - [\phi_{axi},w_{pert}]-[\phi_{pert},w_{axi}] &= 0\\
        \Delta^2 \phi_{pert} + 12 (1-\nu^2)[w_{axi},w_{pert}] &= 0
    \end{split}
\end{equation}

These equations are solved with the same boundary conditions (\ref{eq:bc1}), (\ref{eq:bcsup}) than the full problem, integrating the ODE with \texttt{bvp4c}, and combining this resolution with a root-finding algorithm to determine for each mode $n$ the critical value of the indentation such that $\lambda=0$ (marginal stability).
Since there is an additional parameter $(\lambda)$ in the problem, there is also a supplementary boundary condition that we implement as $w_{pert}(r=1)=1$ which reflects the fact that, at linear order, the amplitude of the buckling mode is indeterminate.

\begin{figure}[!h]
    \centering
    \includegraphics[width=0.45\textwidth]{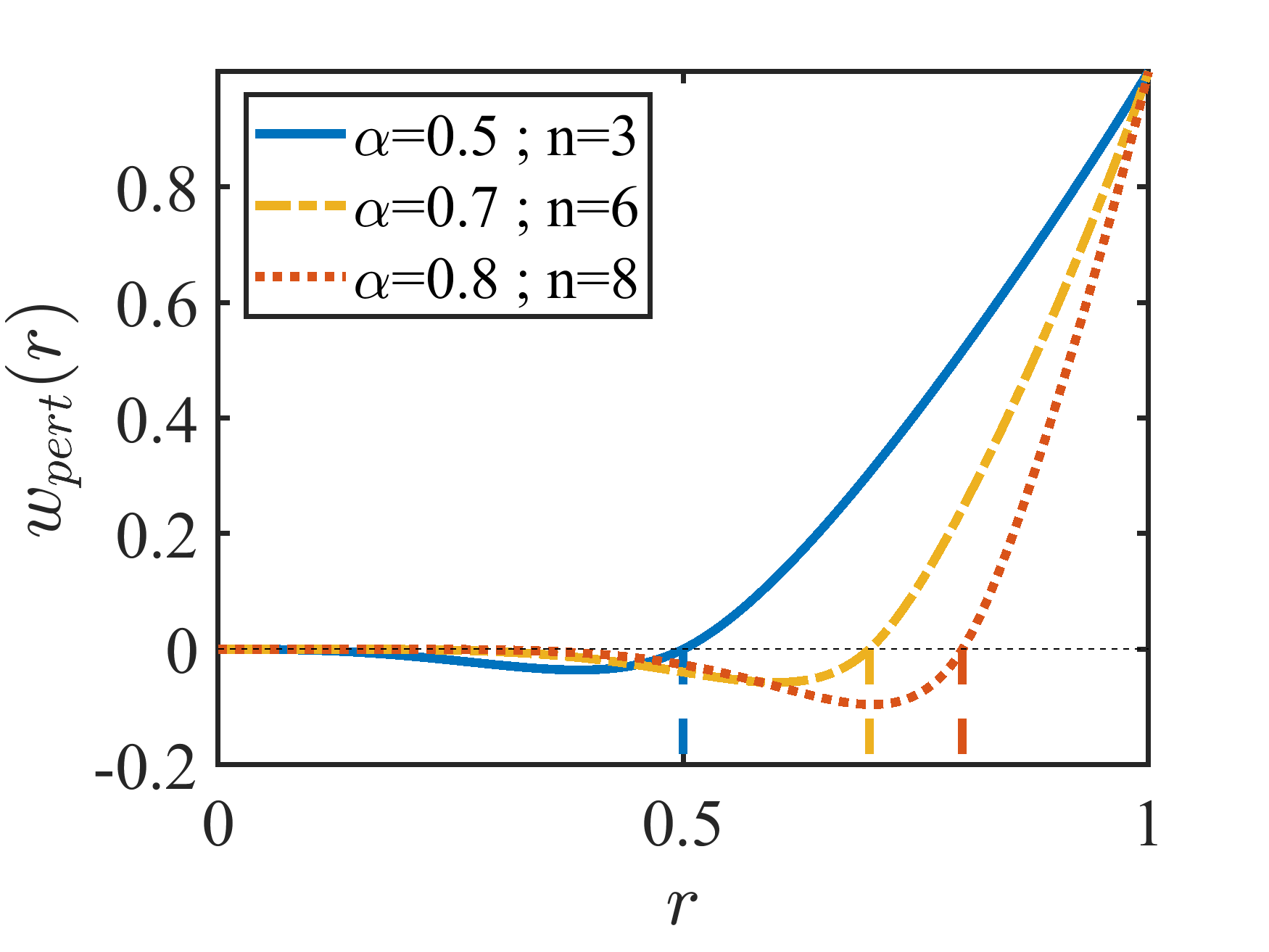}
    \caption{Eigenfunctions $w_{pert}(r)$ computed by linear stability analysis. Location of the supporting ring is indicated by vertical dashed lines.}
    \label{fig:eigenfunctions}
\end{figure}

A few buckling modes $w_{pert}(r)$ are shown in figure \ref{fig:eigenfunctions}. We emphasize that the amplitude of the modes is irrelevant at linear order since it cannot be computed by the stability analysis : however, the spatial structure reflects what can be expected in a weakly nonlinear regime where the system can be described as a combination of a few linearly unstable modes. The spatial structure depends on the aspect ratio and on the mode number, but only weakly on the dimensionless indentation : for a fixed set of $(\alpha,n)$ parameters, changing $\delta$ mostly changes the eigenvalue $\lambda$. For all mode numbers and aspect ratios, the eigenfunction is maximal at the outer edge, and presents a secondary oscillation of opposite sign in the inner region, with an amplitude much lower that the deflection of the outer edge. The fact that the deviation from the axisymmetric state extends into the inner part of the disk, is consistent with the fact that the orthoradial stresses are compressive well into the inner region. However, the amplitude of this secondary oscillation is too small to be experimentally measurable with our setup.

\section{Results}

\subsection{Azimuthal stability diagram}

\begin{figure}[!h]
    \centering
    \includegraphics[scale=0.4]{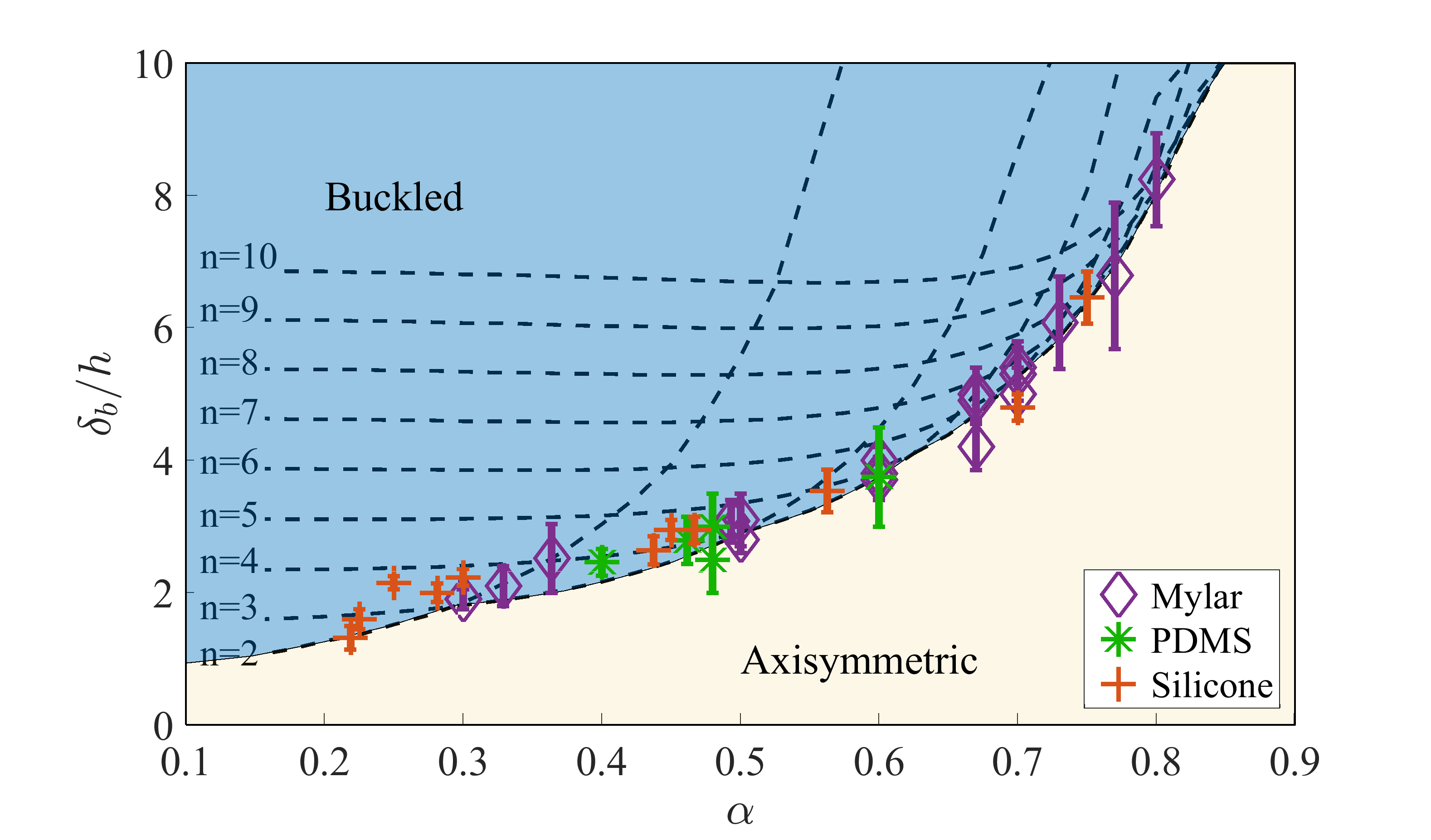}
    \caption{Critical dimensionless indentation for axisymmetric-buckled transition as a function of aspect ratio.}
    \label{fig:diagStabModes}
\end{figure}

Using the method described in section \ref{sec:stability_method}, we obtain the curves of marginal stability for each mode $n$ (changing $n$ corresponds to change the azimuthal wavenumber). For a given aspect ratio, the curve represent the minimal value of the indentation required for the system to be unstable with respect to azimuthal wrinkling. These curves are almost flat for small aspect ratios, and then increase sharply when the aspect ratio tends to 1. The most unstable wavenumber depends on the aspect ratio : e.g. for $\alpha \in [0,0.3]$ the mode $n=2$ is the most unstable, while for $\alpha \in [0.3,0.5]$ the most unstable is $n=3$. As $\alpha$ increases, the most unstable mode corresponds to an increasing $n$, but the regions where only one mode is linearly unstable gets narrower. Thus, in the limit of small overhang of the disk over the support, several different modes can be observed in practice, even close to the boundary: this is consistent with our experiments where, for large $\alpha$, it is generally possible to observe multi-stability close to the wrinkling threshold by manually forcing the disk to adopt a different wavenumber, while for small $\alpha$, only the large wavelength modes $n=2,3$ are observed close to threshold. Note that multistability of the system is also observed for $\alpha \rightarrow 0$ for large indentations (e.g., for $\alpha=0.25, \delta/h = 5$, modes $n=2,3,...7$ are all linearly unstable). We compare these curves of marginal stability with the experimental data obtained following the method described in section \ref{sec:xp_axi_buckling_transition} for different materials. The results are reported in figure \ref{fig:diagStabModes} and shows a good agreement between theory and experimental data. Note that we selected carefully the combination of Young modulus, thickness and size of the sheets such that the gravity parameter is $G < 100$, indicating negligible effects of gravity \cite{Boedec_2021}. 

\subsection{d-cone disappearance threshold}

\begin{figure}[!h]
    \centering
    \includegraphics[scale=0.4]{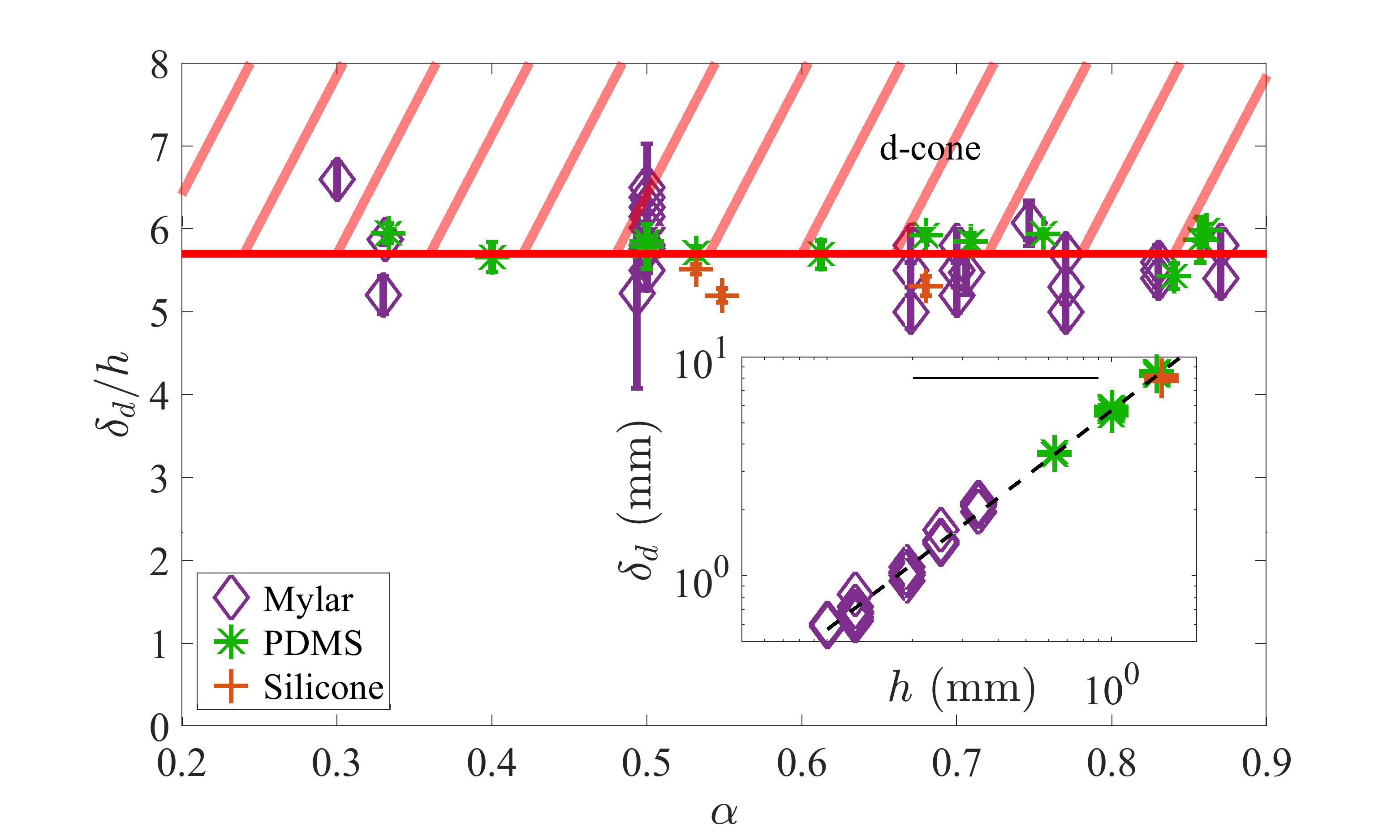}
    \caption{Critical dimensionless indentation for d-cone disappearance as a function of aspect ratio. Inset : critical indentation scales linearly with thickness.}
    \label{fig:disp_dcone}
\end{figure}

As discussed in section \ref{sec:methods-stretch-dcone-transition}, the stretching branch (axisymmetric or wrinkled) and the d-cone branch coexist over a range of indentations. For indentations below a critical value, only the stretching branch is found. We report in figure \ref{fig:disp_dcone} the evolution of this critical value : for aspect ratios ranging from 0.3 to 0.9, the critical dimensionless indentation is essentially constant $\frac{\delta}{h}\approx 5.7$. 

This relationship has been observed for a variety of material with Young modulus varying over three decades and thickness varying over one decade (see inset of figure \ref{fig:disp_dcone}).

An argument for the existence of this threshold can be developed by comparing the energy of the stretching branch to the energy of the d-cone. The energy of the stretching is estimated by considering an axisymmetric state and neglecting all bending contributions. Thus $U_{str}\sim \int_S \sigma_{rr}\epsilon_{r r}\mathrm{d}S \sim Eh \int_{S} \epsilon_{rr}^2 \mathrm{d}S$. We estimate the strain in the inner part of the disk such that $\epsilon_{rr}\sim \left(\dfrac{\delta}{a}\right)^2$ over a surface roughly equal to $a^2$, which leads to

\begin{equation}
    U_{str}\sim E h \left(\frac{\delta}{a}\right)^4 a^2
\end{equation}
On the other hand, a modelling of the d-cone gives its energy as the bending energy of the outer part and a "core" energy. 
Following \cite{Cerda_2005}, the bending energy of the outer part is 
\begin{equation}
    U_{b,out} \sim \left(\frac{\delta}{a}\right)^2 Eh^3 \ln b/r_c
    \label{eq:rc}
\end{equation}
with the core radius $r_c$ given by 
\begin{equation}
    r_c \sim \left(\frac{E h^3}{E h}\right)^{1/6}\left(\frac{\delta}{a}\right)^{-1/3}a^{2/3} \sim \left(\frac{h}{\delta}\right)^{1/3}a
\end{equation}
Note that this scaling yields a core radius larger than the sheet size for $(\delta/h)<\alpha^3$, thus limiting the physically acceptable range of validity of this model to indentations larger than $\alpha^3$.

The core energy is given by a stretching part $U_{s,core}\sim Eh \left(\frac{\delta}{a}\right)^4 \frac{r_c^6}{a^4}$ and a bending part $U_{b,core}\sim Eh^3 \left(\frac{\delta}{a}\right)^2$. Thus, following \cite{Cerda_2005}, the d-cone energy is given by 

\begin{equation}
U_{dc}\sim Eh^3\left(\frac{\delta}{a}\right)^2 \left[1+\ln\left(\frac{b}{a} \frac{\delta^{1/3}}{h^{1/3}}\right)\right]    
\end{equation}
The ratio between the energies of the two branches is :
\begin{equation}
    \frac{U_{dc}}{U_{str}} \sim  \left(\frac{\delta}{h}\right)^{-2}\left[1+\frac{1}{3}\ln\left(\frac{\delta}{h}\right)-\ln \alpha \right]
    \label{eq:ratio-energies}
\end{equation}

For dimensionless indentations in the range $[\alpha^3,\infty[$, the function (\ref{eq:ratio-energies}) is positive and monotonically decreasing from $1/\alpha^{6}$ to 0 : for small indentations it is always more favorable for the system to follow the stretching branch while for large enough indentations d-cone is the least energetic solution. This analysis predicts that the critical indentation for d-cone existence depends linearly on the thickness, which is confirmed by experiments (see inset of figure \ref{fig:disp_dcone}). Note that the argument developed to obtain equation (\ref{eq:ratio-energies}) is only qualitative : to obtain a quantitative scaling of the threshold would require a more precise modeling of the d-cone, especially of the core region. Recent work \cite{Mowitz_2022} may provide a framework to develop this quantitative analysis. It would also be interesting to study the existence of a threshold for e-cones \cite{Efrati_2015,Seffen_2016}.

\section{Summary}
\begin{figure}[!h]
    \includegraphics[scale=0.45]{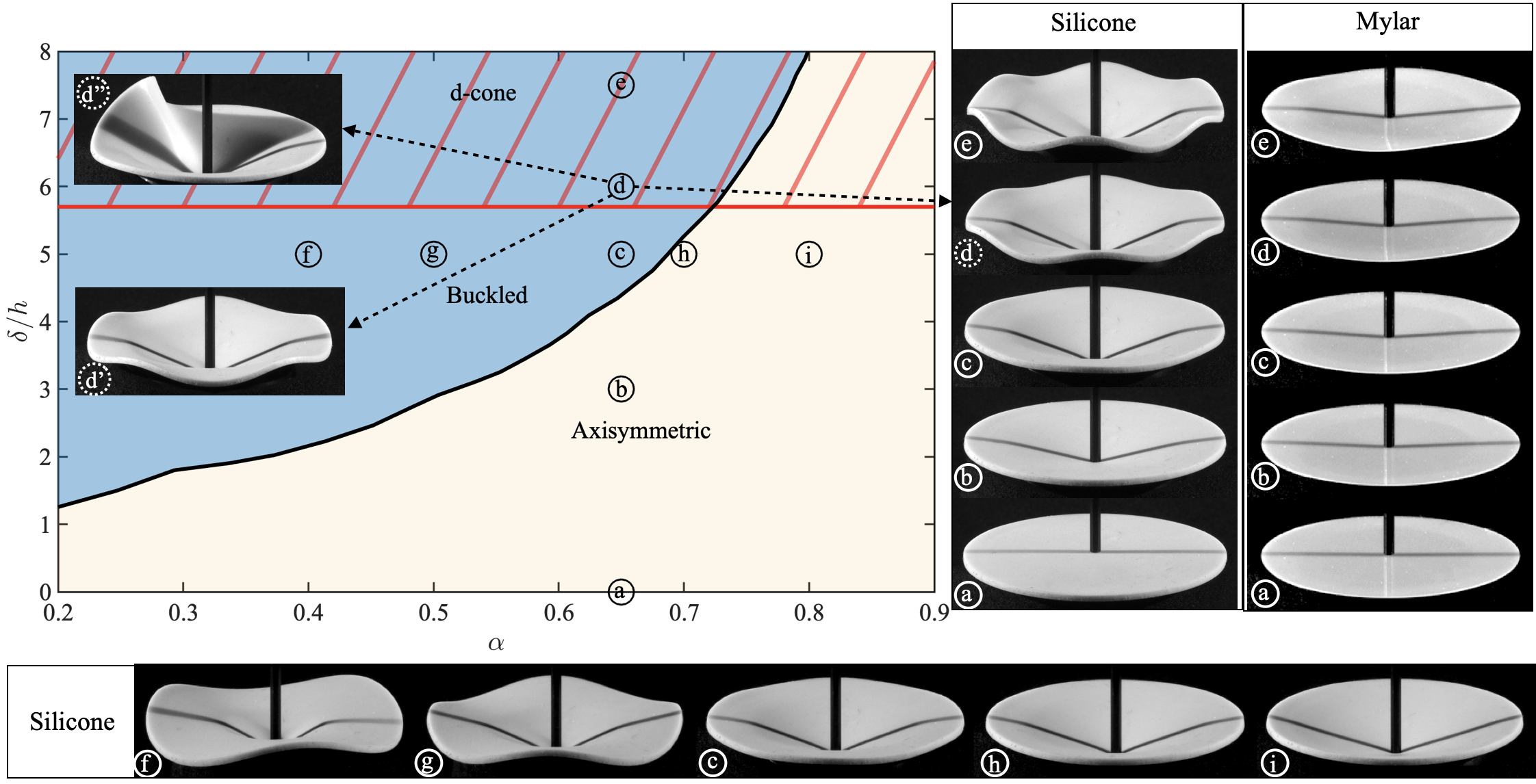}
    \caption{Diagram summarizing the results: several typical shapes are shown for various dimensionless indentations and aspect ratios. Note that beyond stability boundaries, multiple shapes generally coexist, as exemplified at the point of coordinates ($\alpha=0.65,\delta/h=6$) : a mode $n=6$ (picture d) coexists with a mode $n=5$ (picture d') and the d-cone (picture d").  Pictures are taken from an experiment with silicone ($b = 20$ mm, $h = 1$ mm).}
    \label{fig:final diagram}
\end{figure}

We conducted experiments on thin elastic sheets indented over a cylindrical support. Using a combination of visualisation and force-indentation measurements, we obtained an experimental diagram of different behaviors of the system in the limit of negligible effects of gravity: axisymmetric deformations, wrinkled states and d-cone state. The boundaries giving the zone of existence of the different states are obtained as a function of dimensionless indentations and aspect ratio. We also showed that for small $G$ and $\alpha \in[0.3, 0.9]$, the d-cone is observable only for indentations above a critical value of $\delta/h \approx 5.7$.
We propose in figure \ref{fig:final diagram} a diagram summarizing the results. 


A typical sequence of shape transformation (pictures a to e) observed in experiments for silicone and Mylar is shown in figure \ref{fig:final diagram}. The Mylar observations from three-quarter view doesn't allow to reveal the typical behavior that Silicone does. That support the necessity to use deflectometry for Mylar to reveal the instability. In the following we then focus on and describe only the silicone behavior. Starting from a flat state (picture a), the indentation first leads to an axisymmetric state (picture b). After crossing the stability boundary, the wrinkling instabilty appears (picture c): even weak, the undulation is visible by looking closely at the edge of the disk. When the amplitude of the undulations is more pronounced (picture d and e), the global shape is clearer : a mode $n$=6 is observed ($n$=5 for Mylar). 
Note the multi-stability characteristic of the system highlighted by the pictures d, d' and d" : these three pictures, taken at a constant value of parameters, represent buckled modes $n$=6, $n$=5 and d-cone shape respectively. While the previous sequence is for $\alpha$ fixed when the indentation increases, a second one reveals the shape transformation for silicone when $\alpha$ increases for a fixed non-dimensional indentation $\delta/h$ (pictures f to i). This sequence illustrates qualitatively the evolution of the mode number $n$ with aspect ratio. For small aspect ratios we observe small mode number and the mode number increases with aspect ratio : $n=3$ for picture f, $n=4$ for picture g, $n=6$ for picture c.

\section{Discussion and perspectives}

Despite the diversity of behaviors observed in figure \ref{fig:final diagram}, two generic different scenarios may occur during the indentation of the disk, depending on the aspect ratio :  for $\alpha \lesssim 0.7$ starting from zero indentation, the disk first deforms assuming an axisymmetric shape. At a critical indentation $\delta_b$, this shape is unstable with respect to azimuthal wrinkling. At a second critical indentation $\delta_d > \delta_b$, d-cone branch becomes another possible stable solution. For $\alpha \gtrsim 0.7$, starting from zero indentation, the disk first deforms following the axisymmetric branch. At a critical indentation $\delta_d$, the d-cone branch becomes a possible solution (the axisymmetric branch being still stable). At a second critical indentation $\delta_b > \delta_d$, the axisymmetric branch loses stability to azimuthal buckling. In both scenarios, multistability is generally present in the system : far from wrinkling threshold, different azimuthal wavenumbers can be observed. Moreover, close to the d-cone threshold, the stretching branch (axisymmetric or wrinkled) is still locally stable : the system can be switched from d-cone to wrinkled state or the opposite. This is an interesting feature of this system which shall allow to probe further the connection between diffuse states (axisymmetric or wrinkled) and a focused state (d-cone). This duality is present in many elastic systems, in particular in the wrinkle to fold transition \cite{Pocivavsek_2008,Holmes-2010-draping}. The apparition of a localized structure in these systems have been analyzed in terms of nonlinear effects leading to a modulation of the amplitude of the homogeneous pattern \cite{Audoly_2011}. In this regard, the road to a localized pattern is a continuous evolution from a homogeneous pattern whose envelope progressively localize on only one oscillation : whether the transition between stretching and d-cone states belongs to the same class is an attractive perspective that we plan to adress in the future. Other metastable states can also be observed at higher indentations, such as a conical shape with two folds \cite{Cerda_2005}. Another interesting perspective is the effect of gravity: when $G$ parameter is no longer small, we observed in preliminary experiments that a greater variety of behaviors can be observed, such as \textit{g-cones} reported in \cite{Jules_2020}. Gravity also affects the formation of wrinkles, even without central indentation \cite{Boedec_2021} : the diagram \ref{fig:final diagram} can be extended with a third axis, a task that we leave for future works.

\bibliography{biblio}
\bibliographystyle{ieeetr}

\end{document}